\documentclass[fleqn,usenatbib,amsmath,amssymb,newtxtext,newtxmath]{mnras}

\usepackage{times}
\usepackage{mathtools}
\usepackage{graphicx}
\usepackage{dcolumn}
\newcolumntype{d}[1]{D{.}{.}{#1}}
\usepackage{bm}
\usepackage{verbatim} 
\usepackage{cprotect} 
\usepackage{url}
\usepackage{subfigure} 
\usepackage{tikz}
\usepackage{afterpage}
\usepackage{nicefrac}
\usepackage{tabularx}
\newcolumntype{L}{>{\raggedright\arraybackslash}X}
\usepackage{multirow}
\usepackage[english]{babel}
\usepackage{natbib} 
\usepackage{glossaries} 
\usepackage{hyperref}
\usepackage{aas_macros} 
\usepackage{comment}

\graphicspath{{figures/}}

\newcommand{\DocumentID}{P1900059}
 
\usepackage{xcolor} 

\begin{document}

\title[TESLA-X: Sub-threshold Lensed GW Search]{TESLA-X: An effective method to search for sub-threshold lensed gravitational waves with a targeted population model}

\author[Li et al.]{Alvin K. Y. Li$^{1}$\thanks{E-mail: kli7@caltech.edu},
Juno C. L. Chan$^{2}$\thanks{E-mail: chun.lung.chan@nbi.ku.dk},
Heather Fong$^{3,4,5}$\thanks{E-mail: hkyfong@gmail.com},
Aidan H. Y. Chong$^{6}$,
\newauthor{Alan J. Weinstein$^{1}$, 
Jose M. Ezquiaga$^{2}$}
\\
$^{1}$LIGO Laboratory, California Institute of Technology, Pasadena, CA 91125, USA\\
$^{2}$Niels Bohr International Academy, Niels Bohr Institute, Blegdamsvej 17, 2100 Copenhagen, Denmark\\
$^{3}$RESCEU, The University of Tokyo, Tokyo 113-0033, Japan\\
$^{4}$Graduate School of Science, The University of Tokyo, Tokyo 113-0033, Japan\\
$^{5}$University of British Columbia, 2329 West Mall, Vancouver, BC V6T 1Z4, Canada\\
$^{6}$Department of Physics, The Chinese University of Hong Kong, Hong Kong
}
\date{Accepted XXX. Received YYY; in original form ZZZ}
\maketitle
\begin{abstract}
	Strong gravitational lensing can produce copies of gravitational-wave signals from the same source with the same waveform morphologies but different amplitudes and arrival times. Some of these strongly-lensed gravitational-wave signals can be demagnified and become sub-threshold. We present TESLA-X, an enhanced approach to the original GstLAL-based TargetEd Subthreshold Lensing seArch (TESLA) method, for improving the detection efficiency of these potential sub-threshold lensed signals. TESLA-X utilizes lensed injections to generate a targeted population model and a targeted template bank. We compare the performance of a full template bank search, TESLA, and TESLA-X methods via a simulation campaign, and demonstrate the performance of TESLA-X in recovering lensed injections, particularly targeting a mock event. Our results show that the TESLA-X method achieves a maximum of $\sim 10\%$ higher search sensitivity compared to the TESLA method within the sub-threshold regime, presenting a step towards detecting the first lensed gravitational wave. TESLA-X will be employed for the LIGO-Virgo-KAGRA's collaboration-wide analysis to search for lensing signatures in the fourth observing run.
\end{abstract}
\section{Introduction}\label{Section: Introduction}
Gravitational lensing refers to the phenomenon in which properties of propagating waves are influenced by the distortion of spacetime caused by massive intervening objects. The effects of gravitational lensing depends on the mass properties of the lens~\citep{Ohanian:1974ys,Thorne:1982cv,Deguchi:1986zz,Wang:1996as,Nakamura:1997sw,Takahashi:2003ix}. For massive lenses significantly larger than the gravitational-wave wavelength, we can assume geometric optics, which results in repeated signals with relative (de-)magnifications, time delays, and phase shifts~\citep{Wang:1996as,Dai:2017huk,Smith:2017mqu,Smith:2018gle,Smith:2019dis,Li:2018prc,Ng:2017yiu,Oguri:2018muv,Robertson:2020mfh,Ryczanowski:2020mlt,Ezquiaga:2020gdt,Janquart:2021qov,Ezquiaga:2023xfe,Lo:2021nae,Barsode:2024zwv,Chan:2024qmb}. Lenses with sizes comparable to gravitational-wave wavelengths can produce beating patterns through frequency-dependent wave optics effects, formally known as microlensing~\citep{Deguchi:1986zz,Nakamura:1997sw,Takahashi:2003ix,Cao:2014oaa,Christian:2018vsi,Dai:2018enj,Lai:2018rto,Diego:2019lcd,Cheung:2020okf,Cremonese:2021puh,Mishra:2021abc,Meena:2022unp,Caliskan:2022hbu,Caliskan:2023zqm,Chan:2025wgz}.\\
The search for gravitationally-lensed gravitational waves is driven by various scientific objectives, such as testing general relativity~\citep{Baker:2016reh,Collett:2016dey,Ezquiaga:2020dao,Ezquiaga:2020gdt,Goyal:2023uvm,Narayan:2024rat}, improving source localization accuracy~\citep{Hannuksela:2020xor, Yu:2020agu,Vujeva:2024scq,Vujeva:2025kko}, characterizing lens models~\citep{Lai:2018rto, Diego:2019lcd, Oguri:2020ldf, Liu:2023ikc}, and performing precision cosmology~\citep{Sereno:2011ty,Liao:2017ioi,Cao:2019kgn,Hannuksela:2020xor,Farah:2025ews} and statistical cosmology \citep{Xu:2021bfn}. \\
The LIGO-Virgo-KAGRA (LVK) collaboration has conducted various analyses searching for lensing signatures in the event catalogs of LVK's observing runs O1-O2~\citep{Hannuksela:2019kle} and O3~\citep{LIGOScientific:2021izm,LIGOScientific:2023bwz}, but no convincing evidence of lensing signatures has been found. Additionally, other analyses have concluded that there are no confident lensing signatures in the detected gravitational-wave events~\citep{Li:2019osa,McIsaac:2019use,Dai:2020tpj,Liu:2020par,Ezquiaga:2023xfe}.
None the less, \citet{Wierda:2021upe} predicted that the current LVK detector network at its design sensitivity can reach a lensing detection rate of $1.9$ events per year. 
Furthermore, \citet{Wang:2021kzt} projected that in the era of the Cosmic Explorer (CE), the annual rate of strongly lensed gravitational-wave detections could reach as high as $184.7$.

Strongly lensed gravitational waves will likely originate from sources living in the higher redshift universe. Without being lensed, these signals will only be barely detectable or undetectable. Under strong lensing, copies of gravitational waves from these high redshift sources will be produced; some may be magnified and are detected as super-threshold gravitational waves, but some may be potentially de-magnified with a reduced signal-to-noise (SNR) ratio below the detection threshold. These signals, formally known as ``sub-threshold'' signals, are of particular interest, as a significant proportion of strongly-lensed gravitational-wave events falls into this category~\citep{Wierda:2021upe}. Retrieving these possible sub-threshold lensed counterparts is a crucial step in boosting confidence for the first detection of strongly-lensed gravitational waves.\\
Two targeted search methods~\citep{Li:2019osa, McIsaac:2019use} have been developed
and employed to recover potential sub-threshold lensed counterparts to known
super-threshold gravitational waves. The GstLAL-based TargetEd Subthreshold Lensing 
seArch (TESLA) described in~\citet{Li:2019osa} strategically reduces the search parameter 
space to look for possible sub-threshold lensed counterparts. However, as we will explain
in this paper, the traditional TESLA method can potentially lose signals due
to an excessive loss in signal-to-noise (SNR) ratio due to how the targeted
template bank is constructed.\\
Additionally, TESLA's efficiency in finding potential sub-threshold lensed
gravitational waves can be improved if a targeted population model
built based on information from the target superthreshold event is used.
A proper targeted population model in the ranking statistic can further
reduce the influence of the noise background. Detailed information on the
TESLA method can be found in \citep{Li:2019osa}.\\
In this paper, we present a novel method, TESLA-X, to search for
strongly-lensed sub-threshold gravitational-wave signals. TESLA-X employs
a targeted population model to enhance the search sensitivity for potential
lensed sub-threshold signals, and it addresses the excessive SNR loss issue
by utilizing a new targeted template bank construction method.\\
This paper is organized as follows:
Section \ref{Section: Lensing} provides a brief review of strong lensing of gravitational waves. 
Section \ref{Section: sub-threshold} outlines the workflow of the traditional TESLA method.
In Section \ref{Section: TESLA-X}, we give an overview of how the population prior term in the likelihood ratio calculations
for triggers from the GstLAL search pipeline is calculated, and detail the TESLA-X methodology.
Section \ref{Section: MDC} presents a simulation campaign to evaluate the TESLA-X method.
Finally, in Section \ref{Section: Conclusion}, we summarize our findings and discuss potential
future work to further enhance the search sensitivity of the TESLA-X pipeline.

\section{Strong lensing of gravitational waves}\label{Section: Lensing}
Massive objects, like galaxies and galaxy clusters, can produce distortion in spacetime. General relativity predicts that when waves emitted from a source pass by these massive objects (also known as \emph{gravitational lenses}), they will be deflected due to the curvature of spacetime. This effect is known as \emph{gravitational lensing}, which is universal according to the equivalence principle, i.e. it affects gravitational waves in the exact same way as electromagnetic waves (EM waves). Depending on the spatial extent of the potential well of gravitational lenses and the scales of wavelength, the effect of gravitational lensing can be vastly different. In fact, gravitational lensing can be sub-classified into strong lensing, weak lensing, and microlensing. Some literature also distinguishes millilensing as a sub-type of gravitational lensing. In this paper, we focus on \emph{strong lensing} with geometric optics, in which we assume that gravitational waves have a much shorter wavelength than the spatial scale of the gravitational lenses. \\
Under the strong lensing hypothesis, strong gravitational lenses can produce multiple copies of transient gravitational waves from the same source. These repeated signals have identical waveforms (i.e. their underlying intrinsic parameters $\vec{\theta}$, e.g. component masses and spins, are the same), except that (I) their amplitudes may differ by some amplitude scaling factor $\sqrt{\mu_j}$ (Note that the amplitude scaling factor is achromatic under the geometric optics assumption, and the amplification factor can be $>1$ (magnified) or $\leq 1$ (de-magnified).), (II) they are separated by a relative arrival time delay $\Delta t_j$ with respect to the not-lensed signal's coalescence time, and (III) they have an additional Morse phase factor $\Delta \phi_j$ applied to depending on the type of lensed signals (Note: There are three types of lensed signals, type I, II and III. They correspond to the minimum, saddle point, and maximum time-delay solution to the lens equation, respectively.), given by~\citet{Ezquiaga:2020dao, Dai:2020tpj}
\begin{equation}
	\Delta \phi_j = -\frac{n_j \pi}{2},\,\,n_j = 
	\begin{cases}
		0, \text{Type I lensed signals}\\
		1, \text{Type II lensed signals}\\
		2, \text{Type III lensed signals}
	\end{cases}.
\end{equation}
Mathematically, if we denote the not-lensed gravitational wave in the frequency domain as $\tilde{h}^\text{NL}(f; \vec{\theta}, \Delta t_j = 0)$, then the $j^\text{th}$ strongly-lensed counterpart waveform $\tilde{h}_j$ is
\begin{equation}
	\tilde{h}_j (f; \vec{\theta}, \mu_j, \Delta t_j, \Delta \phi_j) = \sqrt{| \mu_j |}\tilde{h}^\mathrm{NL} (f; \vec{\theta}, \Delta t_j) e^{\left(i\,\mathrm{sign}(f) \Delta \phi_j\right)}.
\end{equation}
Since the travelling paths of these repeated signals have been altered, these signals will appear to be originating from sources at different sky locations. However, the deviation in sky location due to strong lensing is in order of arc-seconds. This is negligible compared to the uncertainty in sky localization for gravitational waves, which is in the order of degrees~\citep{KAGRA:2013rdx}. Therefore, for the rest of this paper we will assume multiple strongly-lensed gravitational-wave signals from the same source to have the same sky location.

\section{The search for sub-threshold strongly-lensed gravitational waves: Objectives and current approach}\label{Section: sub-threshold}
\subsection{Current effort to search for strongly-lensed sub-threshold lensed counterparts to super-threshold gravitational waves}

When searching for gravitational waves in general, signals with sufficiently high
amplitudes that can be identified as gravitational waves from the noisy data are
known as \emph{super-threshold}. In contrast, weaker gravitational waves
not distinguishable from noise are known as \emph{sub-threshold}. As discussed in
Section \ref{Section: Lensing}, strong lensing can produce multiple copies
of gravitational waves coming from the same source with different amplification
factors $\sqrt{\mu_j}$. $\sqrt{\mu_j}$ can take on values smaller than $1$,
i.e. the repeated gravitational-wave signals are \emph{de-magnified} compared to
the not-lensed gravitational-wave signal and may become sub-threshold. It is then
natural to ask the question: If we assume that one of the identified super-threshold
gravitational waves detected so far is strongly lensed, can we find its sub-threshold
strongly lensed counterparts, if they exist?\\
There have been ongoing efforts to employ \emph{matched-filtering based search pipelines}
with modifications~\citep{Li:2019osa, McIsaac:2019use} to search for strongly-lensed
sub-threshold lensed counterparts to identified super-threshold gravitational waves
from LVK's first three observing runs (O1, O2 and O3)~\citep{Hannuksela:2019kle,LIGOScientific:2021izm,LIGOScientific:2023bwz}.
In this paper, we focus on the \emph{TargetEd subthreshold Lensing seArch (TESLA)}
method \citep{Li:2019osa}, which is built based on the GstLAL search
pipeline \citep{2017PhRvD..95d2001M,Sachdev:2019vvd,Hanna:2019ezx,Allen:2005fk,2021SoftX..1400680C,Godwin:2020weu,Chan:2020fip,2021PhRvD.103h4047M,2020arXiv201102457M}.
The following sub-section gives a brief introduction to the traditional TESLA method.

\subsection{The traditional TESLA search method}
Traditional TESLA is built based on the matched-filtering-based pipeline GstLAL.
A detailed introduction to the GstLAL search pipeline can be found
in~\citep{Li:2019osa,2017PhRvD..95d2001M,2021SoftX..1400680C}. In a general search
for gravitational waves from compact binary coalescences, the goal is to detect
signals that span the entire parameter space to which the detector network's
frequency bandwidth is sensitive. Hence, a large template bank comprising over
a million template waveforms is constructed and used. A large number of candidates
(also known as triggers) is generated, regardless of whether they are noise or signals.
The more templates we use in a search, the larger the trials factor,
and hence, the larger the noise background will become.\\
Gravitational waves with weaker amplitudes, for instance, strongly-lensed sub-threshold
lensed counterparts, can be buried in the large noise background. Hence, it is necessary
to tactically reduce the noise background while keeping the targeted foreground constant to uncover possible weaker gravitational waves.\\
Assuming a super-threshold gravitational wave is strongly lensed, we want to search for
its possible sub-threshold lensed counterparts. In section \ref{Section: Lensing},
we explained that strongly-lensed gravitational waves from the same source should have
identical intrinsic parameters (e.g. component masses and spins), and hence identical
waveforms. For each super-threshold event (target), we obtain the posterior probability
distribution that gives the best estimates of the source parameters of the target using
the Bayesian parameter estimation analyses described in~\citep{Veitch:2014wba, Ashton:2018jfp, Romero-Shaw:2020owr, Ashton:2021anp}.
We can then narrow down the parameter search space for possible strongly-lensed sub-threshold
counterparts to regions consistent with the target's posterior parameter space.
However, as explained in \citep{Li:2019osa}, the posterior parameter space of the target itself
is insufficient because of noise fluctuations in the data. Non-Gaussianity in the data can
cause false signals (especially those that have weaker amplitudes and are hence sub-threshold),
to be registered by templates that have very different parameters than those of the posterior
samples for the target.\\
TESLA accounts for both the signal sub-space and noise fluctuations in the data when searching
for possible strongly-lensed sub-threshold counterparts to super-threshold gravitational waves.
Here, we outline the steps for running a TESLA analysis. For each super-threshold gravitational
event (target), we start with the posterior samples ordered in decreasing order of log-likelihood
from the parameter estimation for the target. Since the amplitudes of possible strongly-lensed
sub-threshold gravitational waves are lower, they will be registered as triggers in the search
pipeline with lower signal-to-noise ratio $\rho$ (SNR). For each posterior sample, we generate
$1$ injection with the exact same parameters (including luminosity distance to the source $D_L$
and sky location, i.e. right ascension $\alpha$ and declination $\delta$) as the sample, and $9$
extra injections with $D_L$ increased (and hence optimal SNR $\rho_\mathrm{opt}$ decreased) to
mimic the de-magnifying effect of strong lensing on the amplitudes of possible sub-threshold
lensed counterparts, according to the inverse proportionality of $D_L$ and the optimal SNR
$\rho_\mathrm{opt}$:
\begin{equation}
\label{Equation: TESLA-X-rho-distance}
        D_L \propto \frac{1}{\rho_\mathrm{opt}}.
\end{equation}
We require the latter $9$ injections to have detector SNR $\rho_\mathrm{det} \geq 4$ in at
least one detector as a constraint from the GstLAL search pipeline (Note: We do not lower 
the single detector SNR threshold in the GstLAL search pipeline since this will lead to an 
exponential increase in the number of both signal and noise triggers. 
As future work, we will explore the possibility of further lowering the single detector 
SNR threshold.). The set of lensed injections represents the possible sub-threshold lensed counterparts to the
target we are searching for. They are then injected into real data, and we perform an
injection campaign using GstLAL with a full template bank to recover them. After the injection
campaign, templates that can find these injections (Note: An injection is ``found'' if the FAR 
of the associated trigger is $\leq$ $1$ in $30$ days.) are kept to construct a targeted template 
bank to search for possible strongly-lensed sub-threshold counterparts to the target. 
The end product of TESLA is a ranked list of possible lensed counterparts
to the target in increasing order of FARs. Readers are reminded that the FARs (or other ranking
statistics) reported in the search's ranked list do not indicate how likely the signal candidate
is a lensed counterpart to the target, but only a priority label for follow-up analysis.\\
For a detailed explanation of the TESLA method, please refer to \citep{Li:2019osa}.

\section{The upgraded TESLA-X search method}\label{Section: TESLA-X}
\subsection{The missing pieces in the traditional TESLA method: Non-optimal template bank and population model}

We have shown in \citep{Li:2019osa} that the traditional TESLA method succeeds in improving
the search sensitivity to possible strongly-lensed sub-threshold gravitational waves.
However, the whole picture still needs to include two puzzle pieces.\\
First, in the original full template bank, the parameter ranges are informed by the masses and
spins of detectable sources (BNS, NSBH, BBH), and the templates are distributed with a density
that allows a maximum loss of $3\%$ SNR when a gravitational-wave signal is recovered (compared
to the optimal SNR (Note: The optimal SNR is the SNR when the signal is cross-correlated with a 
template with the exact same parameters as the signal) \citep{2021PhRvD.103h4047M}.
However, when a reduced template bank is generated using traditional TESLA, we discard templates
that did not recover lensed injections for the target super-threshold gravitational waves. This
breaks the original optimality of the full template bank, caused by the finite number of lensed
injections that are used to generate the reduced template bank, and strongly-lensed sub-threshold
signals whose optimal detector SNR is near threshold (i.e. around $\rho_\mathrm{opt}\approx 4.12$)
will not be detected. Therefore, it is vital that the optimality of the reduced template bank
must be conserved to maximize our search sensitivity towards possible strongly-lensed sub-threshold
gravitational waves.\\
Second, a population model was not set in traditional TESLA. In the analysis, the absence of a
population model carries an implicit assumption that every template in the template bank has an equal
probability of recovering a possible sub-threshold counterpart. Since our goal is to detect sub-threshold
counterparts whose intrinsic parameters are identical to a super-threshold signal, implementing a
population model centered around the signal's recovered parameters will increase the pipeline's
sensitivity in detecting signals in that region. The population model should also be dependent
on SNR to account for sub-threshold signals that are coloured by noise and, thus, could be recovered
by a template that is different from that recovered the super-threshold signal.

\subsection{The likelihood ratio in the GstLAL search pipeline}
%\FIXME{Heather: omit the log part in the text and equation just to keep things simpler - log likelihood is what we use in practice just because it's more convenient but the math is the same}
To rank candidates, GstLAL assigns each trigger a likelihood ratio as its ranking statistic, defined
as~\citep{Sachdev:2019vvd,2021SoftX..1400680C,2017PhRvD..95d2001M}:
\begin{equation}
        \mathcal{L} = \frac{ P(\vec{D}_H, \vec{O}, \vec{\rho}, \vec{\xi}^2, \left[\Delta \vec{t}, \Delta \vec{\phi} \right] | \vec{\theta}, \mathrm{signal}) }{P(\vec{D}_H, \vec{O}, \vec{\rho}, \vec{\xi}^2, \left[\Delta \vec{t}, \Delta \vec{\phi} \right] | \vec{\theta}, \mathrm{noise})} \cdot \mathcal{L}(\theta),
\label{eqn:TESLA-X-likelihood_ratio}
\end{equation}
where the first term, $P(...|\vec{\theta}, \mathrm{signal})/P(...|\vec{\theta}, \mathrm{noise})$, is
the ratio of the probability of obtaining the candidate event under the signal model to that under
the noise model. It depends on
(1) the detectors $\vec{O}$ that are participating at the time of the candidate event,
(2) the horizon distances of the participating detectors $\vec{D}_H$,
(3) the detector SNRs $\vec{\rho}$ and
(4) The candidate event's auto-correlation-based signal consistency test values at each detector $\vec{\xi}^2$.
If a candidate event is found in coincidence at multiple detectors (i.e. the event is a \emph{coincident}
event), $\mathcal{L}$ will also include
(5) the relative time delays $\Delta \vec{t}$ and
(6) the relative phase delays $\Delta \vec{\phi}$ of the candidate event between participating detectors (Note: These relative time delays and phase delays should not be confused with the relative time delays and phase delays coming from strong lensing.).\\
The second term in Equation \ref{eqn:TESLA-X-likelihood_ratio},
\begin{equation}
        \mathcal{L}(\theta) = \frac{ P(\vec{\theta} | \mathrm{signal} ) }{ P(\vec{\theta} | \mathrm{noise}) },
\end{equation}
is the ratio of the probability that a trigger is recovered by a template with parameters $\vec{\theta}$,
given the trigger is an astrophysical signal or noise. Because this ratio does not depend on the data
observed by the detectors and depends on input parameters given before performing the search,
it has been written as a separate term.\\
The method of calculating the numerator and denominator are described in detail in \citep{Heather} and \citep{2015arXiv150404632C},
respectively. While the derivations are not repeated here, we wish to highlight that information about
the population model is enfolded into the numerator; therefore, it is $P(\vec{\theta} | \mathrm{signal} )$
that allows the pipeline to perform targeted source searches. The following section describes the changes
made to improve the strongly-lensed gravitational-wave search.

\subsection{A better reduced template bank, and a lensing-based targeted population model}
Traditional TESLA aims to reduce the noise background effectively to uncover possible strongly-lensed
counterparts to the target super-threshold gravitational wave. We reduce the number of templates,
and hence trials factors, by keeping only templates that can cover the parameter space where possible
lensed counterparts may live in. Through an injection campaign, traditional TESLA locates a subset of
templates from the full template bank to construct a reduced template bank to search for possible lensed
counterparts to a given target super-threshold gravitational wave. However, as explained at the beginning
of \ref{Section: TESLA-X}, there are two problems in traditional TESLA, both of which lead to a decrease in
the search sensitivity of TESLA.\\
We now propose two improvements to the TESLA method that will resolve these problems:
(1) constructing a better reduced template bank, and
(2) generating a targeted population model.\\
We recall that the sole purpose of the injection campaign is to determine the search parameter subspace
that we should focus on to look for possible strongly-lensed sub-threshold counterparts to the target
super-threshold gravitational waves. The injection campaign is always imperfect because we can only do
a limited number of injections with discretized SNRs, i.e. we cannot cover the full spectrum of SNRs that
possible lensed sub-threshold signals can have. Therefore, the templates that can recover lensed
injections in the injection campaign can only serve as a pointer as to the region of search parameter space
we should be targeting, i.e. a search parameter subspace where possible lensed sub-threshold counterparts
to the target super-threshold gravitational waves may live in. We should hence use the injection
campaign results to \emph{define a boundary} within which templates live should be used collectively to
construct the reduced template bank to search for possible lensed counterparts to the target, regardless
of whether the templates can recover injections from the injection campaign. The proposed
reduced template bank does not discard templates within the narrowed-down search parameter space defined
by the boundary. Hence, it maintains the optimality guaranteed by the original full template bank
within the limited search space. In turn, any possible lensed counterparts that live within the boundary
will be recovered with an SNR loss less than $3\%$, preventing any loss of potential signals due to
excessive SNR loss.\\
Under the strong lensing hypothesis, possible lensed counterparts should have the same parameters as the target super-threshold gravitational waves regardless of being super-threshold or sub-threshold.
Therefore, the population of possible lensed counterparts to the target gravitational waves should have
a higher probability of lying near the signal subspace defined by the posterior parameter
probability distribution of the super-threshold event, and a lower probability
of living far away from that. When we perform the injection campaign in
traditional TESLA, we can obtain both the recovered injections' SNRs and
the corresponding templates that recover the injections. Generally speaking,
sub-threshold signals with lower optimal SNRs are affected by noise
fluctuations more substantially and hence can be recovered by templates with
parameters different from those of the target super-threshold event. That said,
the templates closer to the signal subspace defined by the posterior
parameter probability distribution of the target are associated with recovered
injections with higher SNR values. Therefore, we can make use of the results
from the injection campaign to construct a \emph{Gaussian Kernel Density Estimation}
(KDE) function $\text{KDE}(\vec\gamma)$ using the distribution of the templates that
recovered the lensed injections, with the corresponding injections' recovered SNRs
as weights. The KDE is a function of source parameters $\vec\gamma$, which, to match
the parameters of the template bank, are set to be the masses and spins of the source.
When weighted with the recovered SNRs, regions closer to the signal subspace
that the target super-threshold gravitational wave lies in will give a higher value
for the KDE function, which matches our expectation that lensed counterparts to the
target should have similar parameters as the target. The KDE function can then be
redefined as the probability density function that describes the ``source population''
for our possible strongly-lensed sub-threshold counterparts to the target gravitational
wave. Using this as our population model, we then follow the steps detailed in \citep{Heather}
to calculate $P(\vec{\theta}|\mathrm{signal})$, i.e. the probability that a signal is
recovered by a template with parameters $\vec{\theta}$. Notably, this probability depends on SNR, such that the search does not overly penalize templates far
away from the signal subspace for weak sub-threshold signals. These methods instill a
correct population model to the new TESLA method compared to the incorrect uniform
population model in traditional TESLA, improving the search sensitivity.

\subsection{The flow of the TESLA-X search method}
We now describe in detail the actual flow of the proposed TESLA-X search method.

\begin{figure}
\includegraphics[width=\columnwidth]{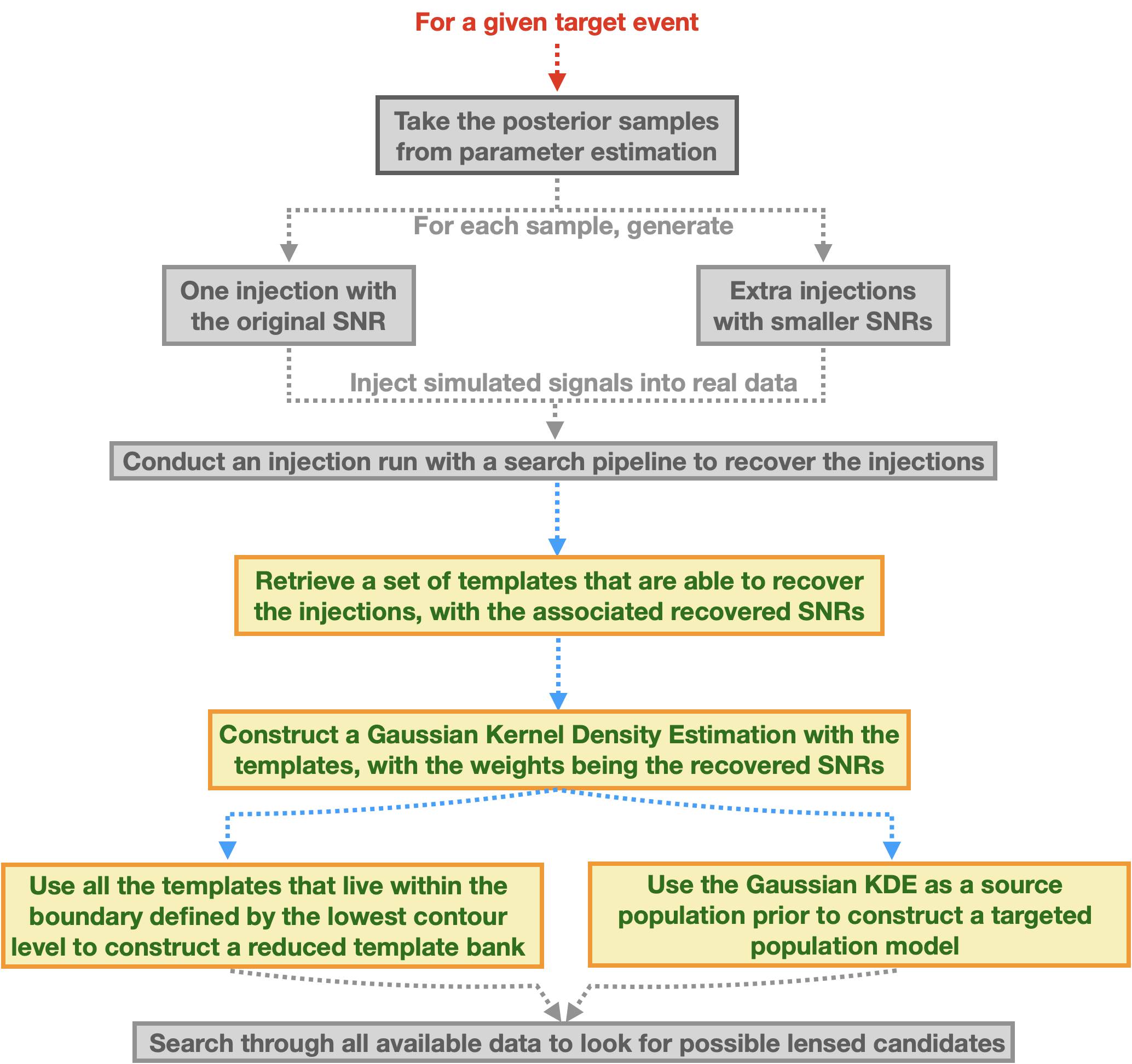}
\caption{\label{Fig: TESLA-X} Workflow of TESLA-X. Elements in grey are the same procedures as in traditional TESLA. Colored elements (color online) are new procedures in TESLA-X.}
\end{figure}

\begin{figure}
\centering
\includegraphics[width=\columnwidth]{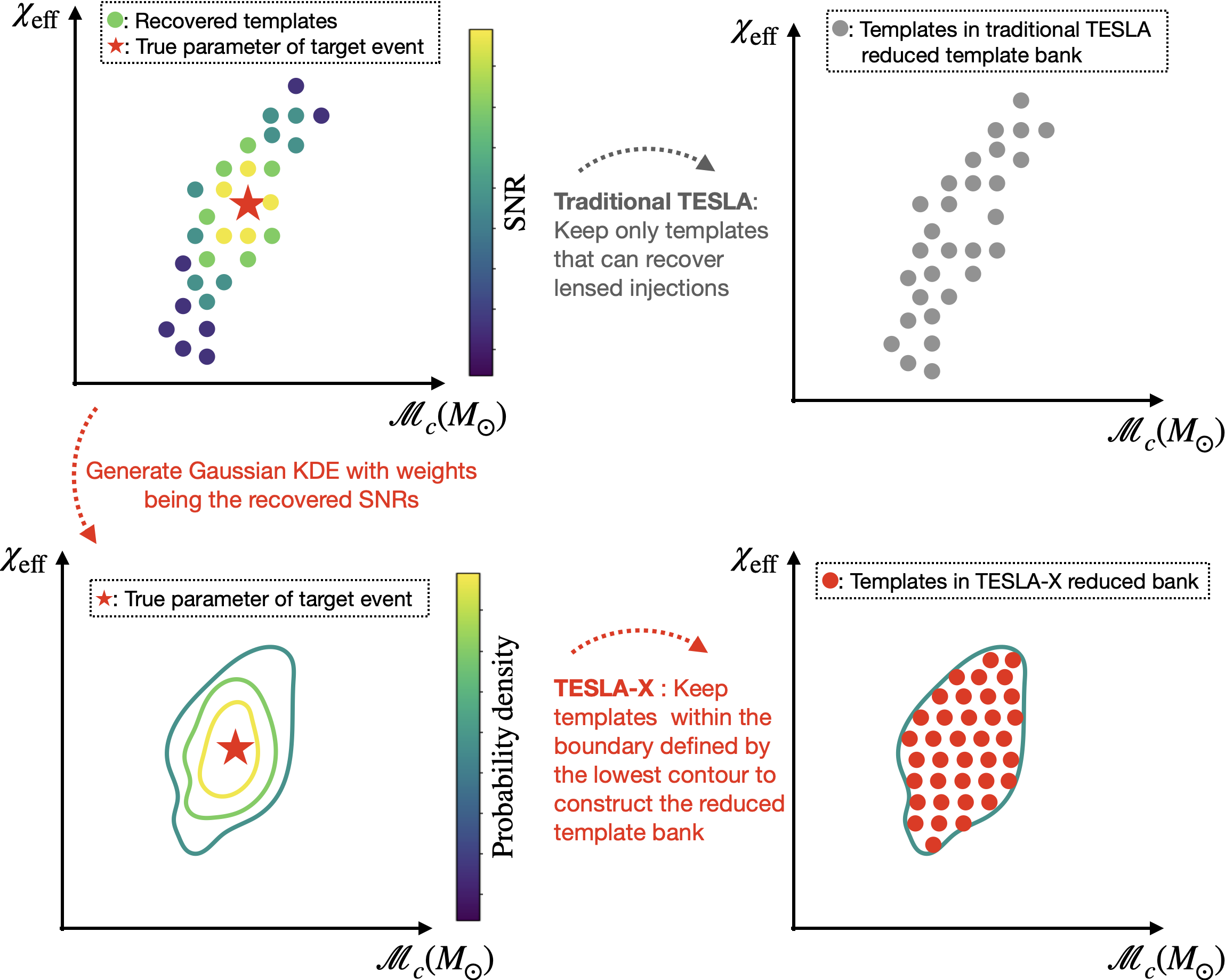}
\caption{
        A cartoon depicting how we construct a reduced template bank from the injection
        results following the Traditional TESLA method (top right) and the proposed
        TESLA-X method (bottom left and right).
}
\label{Fig: TESLA-X-reduced-bank-cartoon}
\index{figures}
\end{figure}

\begin{figure}
\centering
\includegraphics[width=\columnwidth]{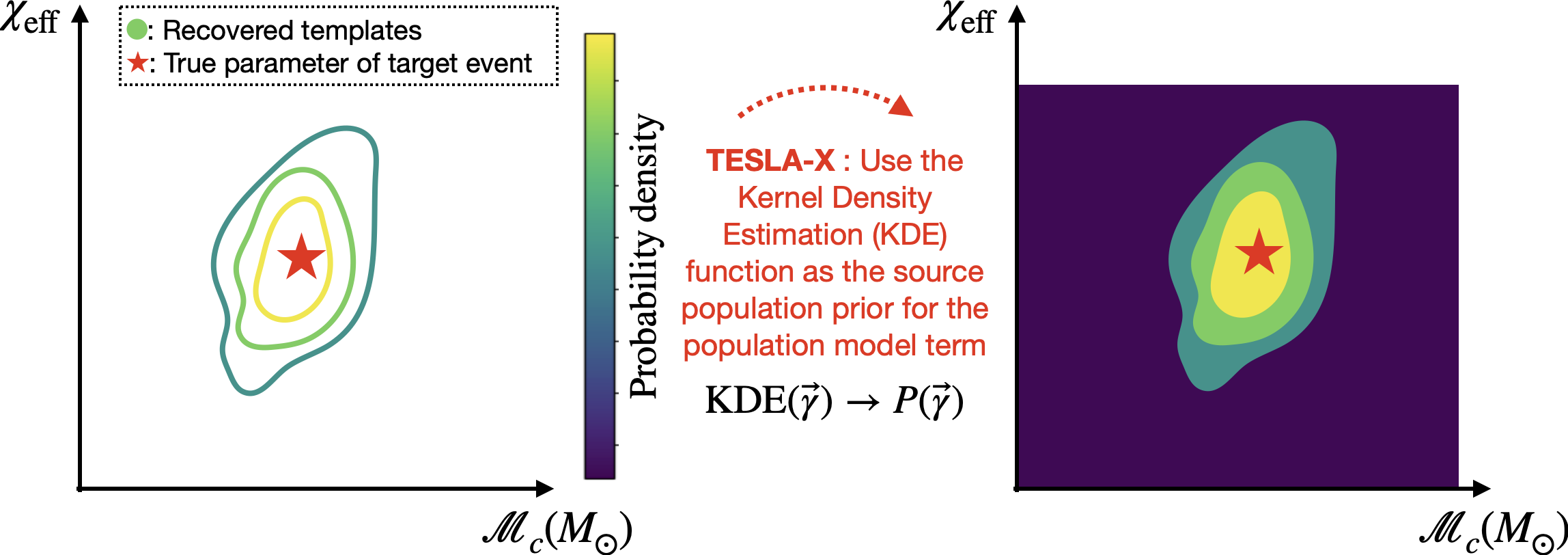}
\caption{
        A cartoon depicting how we use the injection campaign results to construct a targeted
        population model in TESLA-X.
        }
\label{Fig: TESLA-X-mass-model-cartoon}
\index{figures}
\end{figure}

Figure \ref{Fig: TESLA-X} presents the workflow of the TESLA-X method.
We start with a target super-threshold gravitational-wave event found
during a general search with GstLAL using a full template bank. Using the posterior
samples obtained by Bayesian parameter estimation of the target event, we generate,
for each sample, one injection that has the same optimal SNR as the sample, and nine
additional injections with smaller optimal SNRs by increasing their effective distances,
requiring that their SNRs in at least one detector have to be $\geq 4$ to ensure they
can be registered as a trigger during the matched-filtering process in the search.
These injections will represent possible sub-threshold lensed counterparts to the
target super-threshold event. We then inject these simulated lensed injections into
real data, and use the GstLAL search pipeline to try and recover them with a full
template bank. Up to this stage, TESLA-X is still identical to the traditional
TESLA method.\\
After the injection campaign, we now retrieve a set of templates that can
recover the simulated lensed injections, and the associated observed SNRs of the
recovered injections. We can plot the retrieved templates on the chirp mass
$\mathcal{M}_c$ - effective spin $\chi_\text{eff}$ space, with the colors of the
points being the associated recovered injections' observed SNRs $\rho_\text{obs}$.
The top left panel of Figure \ref{Fig: TESLA-X-reduced-bank-cartoon} shows an example
cartoon that plots the templates on the $\mathcal{M}_c\mathrm{-}\chi_\text{eff}$ space,
with the color being the recovered injections' network SNRs.
Traditional TESLA keeps only templates that can recover the lensed injections
and use them to construct the reduced bank, hereby known as ``TESLA BANK'' (top right
panel of Figure \ref{Fig: TESLA-X-reduced-bank-cartoon}). For TESLA-X, we construct
a Gaussian Kernel Density Estimation (KDE) function $f(\vec{\gamma})$ for the templates
that recover the injections, with the corresponding recovered SNRs as the weights, where
$\gamma$ is the set of template parameters we are considering; for example, in
Figure \ref{Fig: TESLA-X-reduced-bank-cartoon} $\vec\gamma$ will be chirp mass $\mathcal{M}_c$
and effective spin $\chi_\text{eff}$. \\
Note that the lensed injections are generated based on the posterior samples
from the parameter estimation of the target event. Spins, admittedly, are not
well measured for gravitational waves. However, recall that the lensed injections'
spin distribution follows the distribution of the parameter estimation
result for the target event. Suppose the spin measurement for the target event is poor. In that case,
the parameter estimation result will essentially return the uninformed, uniform
prior for spins, meaning that the lensed injections' spins will follow a uniform
distribution. Hence, the injection campaign result is not biased towards
spins. Hence, we argue that there is no harm in using effective spin as a parameter
when generating the Gaussian KDE.\\
Using the KDE function, we can then evaluate the probability density for each template point
$\vec\theta$ in the full template bank. We then use ``matplotlib.pyplot.contour'' with the
default settings of ``matplotlib.ticker.MaxNLocator'' to determine the characteristic contour
levels for the KDE function (see bottom left panel of Figure \ref{Fig: TESLA-X-reduced-bank-cartoon})
for example). Then, using the lowest contour (it is left as future work to better define the lowest contour / least significant contour in terms of confidence interval) as the boundary, TESLA-X keeps \emph{all} the
templates within the boundary and uses them to construct the reduced template bank, hereby
known as ``TESLA-X BANK''. While the TESLA BANK has ``holes" (see, for example, the top right picture in Figure \ref{Fig: TESLA-X-reduced-bank-cartoon}) compared to the original full template bank because we are overly removing templates, the TESLA-X BANK keeps all the templates within a certain boundary, hence recovering the optimality of the original template bank in the region within the boundary. Hence, we will not lose signals due to excessive loss in SNR, and we will expect the TESLA-X BANK to give better search sensitivity.\\
Because the Gaussian KDE function $f(\vec\gamma)$ is essentially a probability density function
(PDF) that describes the expected distribution of possible (sub-threshold) lensed counterparts
to the target super-threshold gravitational wave, we can use it as the astrophysical source
population prior to construct a targeted population model using the same methodology described
in the previous section and \citep{Heather}. Figure \ref{Fig: TESLA-X-mass-model-cartoon}
shows an example cartoon to illustrate the idea.
We have also compared the performance of TESLA-X using (1) the targeted population model suggested above, and (2) an empirical, uniform mass model. We have verified that the search with the suggested targeted population model gives a better performance than the uniform mass model.\\
Once the reduced bank and the targeted population model have been constructed, we again use
GstLAL to search through all possible data with the targeted bank and the targeted population
model to look for potential sub-threshold lensed counterparts to the target super-threshold
event. GstLAL will then output a list of candidate events ranked by their assigned
ranking statistics, including the FARs and the likelihood ratios $\mathcal{L}$. We
have to stress again that the FARs assigned to the candidate events here do not quantify
how likely they are to be lensed counterparts to the target event. Still, they are measures to
distinguish noise triggers (false alarms) from real astrophysical signals, whether or not they
are lensed counterparts to a target event. The ranking statistics here should only be used as
a priority ranking for follow-up analysis to decide how likely each candidate event is a lensed
counterpart to the target event. As future work, we will explore the idea of evaluating an additional lensing likelihood ratio for each trigger that includes additional lensing information
as a separate ranking statistic. At that stage, the lensing likelihood may reveal how likely a trigger is a lensed counterpart to the target event.

\section{Method verification and results} \label{Section: MDC}
In \citep{Li:2019osa}, we carry out a full-scale simulation campaign, where we prepare mock
data with a pair of super-threshold and sub-threshold lensed signals injected, and apply
the TESLA method to recover the two signals. We have proved that the TESLA method can recover sub-threshold lensed signals if they exist. TESLA-X and TESLA
have essentially the same working principle, but the only differences are how we construct
the reduced template bank and the population model. Therefore, we conduct a simulation
campaign solely to compare the search effectiveness and sensitivity towards possible
sub-threshold lensed gravitational-wave signals using TESLA and the proposed TESLA-X method.

\subsection{Mock data generation}
First, we generate a $9.15$-day long data stream with Gaussian noise recolored with O4
characteristic power spectral densities for LIGO Hanford, LIGO Livingston, and Virgo
detector.  The exact data for O4 characteristic power spectral densities for LIGO Hanford, 
LIGO Livingston and Virgo detector can be found at \url{https://dcc.ligo.org/T2200043-v3/public}.
We assume that there is no detector downtime (Note: A detector is considered ``down'' if it is not in observing mode.), and no times are vetoed. We inject a super-threshold gravitational wave generated
using the IMRPhenomXPHMpseudoFourPN waveform~\citep{Pratten:2020ceb} into the mock data.
Details about the source parameters of the injected gravitational wave are listed in
Table \ref{Table: MS220508a}.

\begin{table}
%\begin{ruledtabular}
\begin{tabular}{c|c}
Properties & Injected super-threshold signal \\
\hline
UTC time & May 08 2022 $11:06:00$\\
GPS time & $1336043178.397$ \\
Distance (Mpc) & $2858.18$  \\
\hline
Primary mass $m_1^\text{det}$ & {$70.08 M_\odot$}\\
Secondary mass $m_2^\text{det}$ & {$38.83 M_\odot$}\\
Dimensionless spins & {$\chi_{1x}=0.182$}, {$\chi_{1y}=0.182$}, {$\chi_{1z}=-0.0363$},\\
& {$\chi_{2x}=-0.113$}, {$\chi_{2y}=0.132$}, {$\chi_{2z}=0.116$},\\
Right ascension $\alpha$ & {$2.811$} \\
Declination $\delta$ & {$0.819$} \\
Inclination $\iota$ & {$2.513$}\\
Polarization $\Psi$ & {$1.187$}\\
Waveform & IMRPhenomXPHMpseudoFourPN \\
\end{tabular}
\caption{\label{Table: MS220508a}Information of the injected super-threshold gravitational-wave signal MS220508a in the simulation campaign. All properties reported here are measured in the detector frame.}
\end{table}
In later parts of this paper, we may refer to the super-threshold signal as MS220508a.

\subsection{Performing a general search}
We then use GstLAL to perform a search over the mock data stream following the same settings
used to search for gravitational waves within O3 data in GWTC-3 \citep{ligo_scientific_collaboration_and_virgo_2021_5546663}.
We use the same general template bank as described in \citep{Li:2019osa} for the general search.
As expected, the search recovers MS220508a with the highest ranking statistics (FAR$=2.972\times10^{-35}$ Hz,
rank 1) among all other triggers. We then apply a Bayesian inference library for gravitational-wave
astronomy Bilby \citep{Ashton:2018jfp}
\footnote{
        For reference, we used the default MCMC approach to do the sampling.
}
to perform parameter estimation (PE) for MS220508a, which outputs a set of posterior
samples required to apply the TESLA and TESLA-X search pipeline.

\subsection{Performing an injection campaign}
\begin{figure}
\centering
\includegraphics[width=\columnwidth]{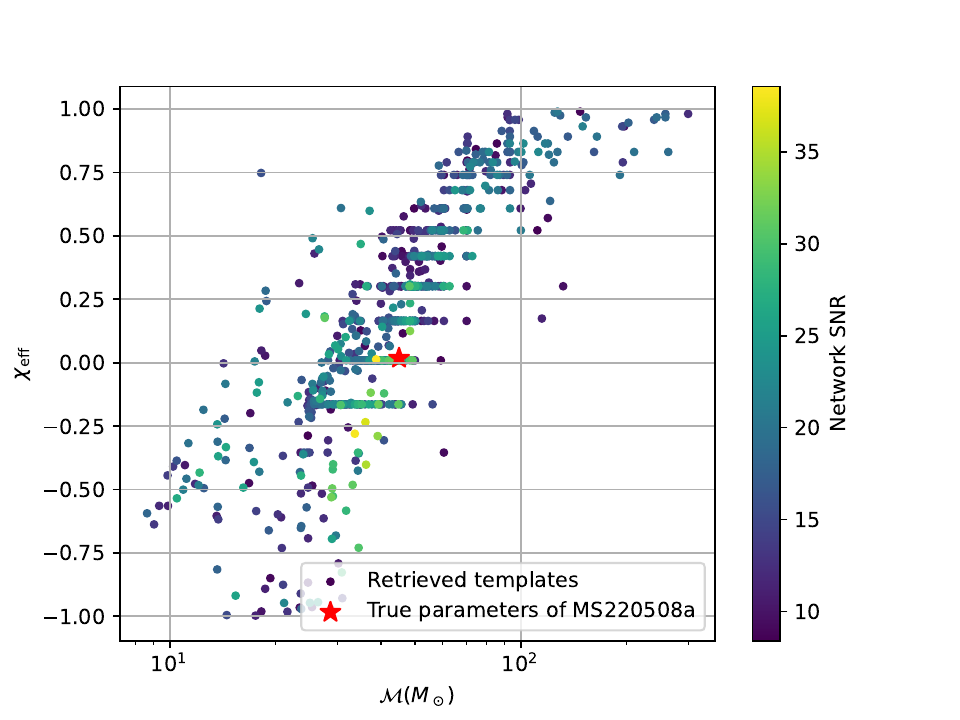}
\caption{
        Rung-up templates by the simulated lensed injections of MS220508a in the injection
        campaign, plotted on the chirp mass $\mathcal{M}_c$ - effective spin $\chi_\text{eff}$ space.
        The colors of the markers represent the network SNRs of the associated injections.
}
\label{Fig: MS220508a recovered templates}
\index{figures}
\end{figure}

Using the posterior samples from the PE for MS220508a, we generated $7815$ simulated lensed
injections following equation \ref{Equation: TESLA-X-rho-distance}. These simulated signals
are injected into the mock data, and we perform another search using GstLAL with the general
template bank to recover them. The results of the injection campaign are then used
to perform the TESLA and TESLA-X analyses.\\
In the injection campaign, the general template bank recovers $6151$ out of $7815$ injections
(See the column ``General'' in Table \ref{Table: TESLA-X Injection recovery}), ringing up a total
of $1008$ templates (Note: The same template may be rung up multiple times by different injections,
and hence the number of rung-up templates can be fewer than the number of found injections.).
Figure \ref{Fig: MS220508a recovered templates} plots the rung-up templates on the chirp mass
$\mathcal{M}_c$ - effective spin $\chi_\text{eff}$ space, with the colors of the markers being
the network SNRs of the associated injections. We can see that the results match our expectations:
Templates closer to the actual parameters of MS220508a are associated with injections
with higher recovered SNRs (i.e. super-threshold). Those that have parameters much different than
the actual parameters of MS220508a are, in general, related to injections with lower
SNRs (i.e. sub-threshold).

\subsection{Generating the TESLA and  TESLA-X reduced template banks}
\begin{figure}
\centering
\includegraphics[width=\columnwidth]{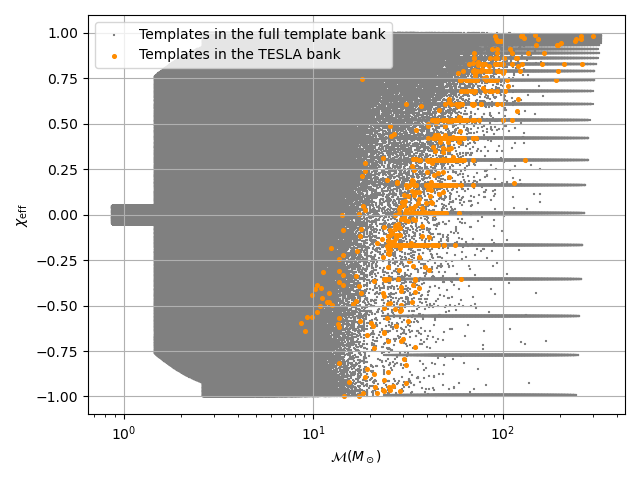}
\caption{
        The templates in the full template bank (in grey) and in the TESLA reduced template
        bank (in orange), plotted in the $\mathcal{M}_c$ - $\chi_\text{eff}$ space.
}
\label{Fig: MS220508a TESLA template bank}
\index{figures}
\end{figure}

\begin{figure}
\centering
\includegraphics[width=\columnwidth]{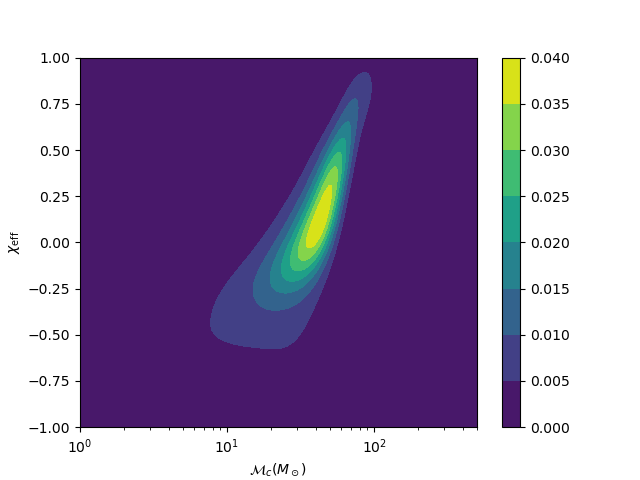}
\caption{
        The contour map of the Gaussian Kernel Density Estimation function obtained for
        MS220508a in the $\mathcal{M}_c$ - $\chi_\text{eff}$ space. The colors represent
        the probability density estimated by the KDE function.
}
\label{Fig: MS220508a Gaussian KDE}
\index{figures}
\end{figure}

\begin{figure}
\centering
\includegraphics[width=\columnwidth]{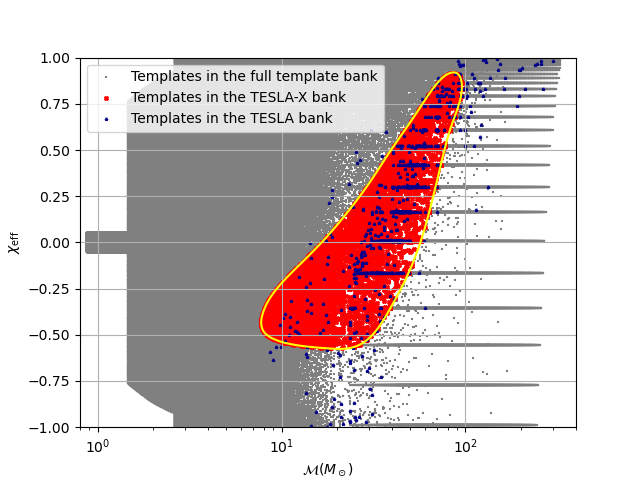}
\caption{
        The templates in the full template bank (in grey) and in the TESLA-X reduced template bank (in red),
        plotted in the $\mathcal{M}_c$ - $\chi_\text{eff}$ space. The TESLA-X template bank is constructed by
        keeping templates that fall within the lowest contour level from the Gaussian KDE function shown in
        Figure \ref{Fig: MS220508a Gaussian KDE}, i.e. $0.005$, plotted as a yellow curve in this
        figure. For easy comparison, we plot the TESLA reduced template bank templates in the
        same figure (in blue).
}
\label{Fig: MS220508a TESLA-X bank}
\index{figures}
\end{figure}

We keep all templates that are rung up during the injection campaign, and use
them to generate a TESLA targeted template bank. Figure \ref{Fig: MS220508a TESLA template bank}
shows the templates in the full template bank (in grey) and those in the TESLA template bank
(in orange).
For TESLA-X, we first use the rung-up templates in the injection campaign to construct a
Gaussian Kernel Density Estimation (KDE) function $\text{KDE}(\vec{\gamma})$ with the
associated injections' network SNRs as the weights. $\gamma$ represents the template
parameters used to label the templates. Here, $\vec\gamma = \{\mathcal{M}_c, \chi_\text{eff}\}$.
Then, we evaluate the probability density at each template point in the full template bank,
and plot the results on a contour map. Figure \ref{Fig: MS220508a Gaussian KDE} shows the
contour map for the Gaussian KDE function estimated for MS220508a.
We then take the lowest contour level (For the case of MS220508a, the lowest contour level
has a value of $0.005$) as a boundary, and use all the templates within this
boundary to construct a TESLA-X targeted template bank. Figure \ref{Fig: MS220508a TESLA-X bank}
shows the templates in the full template bank (in grey) and in the TESLA-X reduced template
bank (in red) in the $\mathcal{M}_c$ - $\chi_\text{eff}$ space. The yellow curve marks the
lowest contour level we obtained for the estimated Gaussian KDE function we showed in
Figure \ref{Fig: MS220508a Gaussian KDE}. There are $22136$ templates in the TESLA-X reduced
template bank.

\subsection{Constructing the targeted population model for the TESLA-X reduced template bank}
Next, we use the estimated KDE function for MS220508a and the TESLA-X reduced template bank
to construct a targeted population model, following the same steps detailed in \ref{Section: TESLA-X}
and \citep{Heather}. Note that for practical reasons, the Gaussian KDE function $\text{KDE}(\vec\gamma)$
is re-calculated with $\vec\gamma$ chosen to be the component masses and spins $m_1, m_2, \chi_1, \chi_2$
instead of $\mathcal{M}_c$ and $\chi_\text{eff}$.

\subsection{Performing a re-filtering to recover the lensed injections with the TESLA and TESLA-X methods}
Finally, we perform two searches, one using the TESLA reduced template bank, and one using
the TESLA-X reduced template bank, together with the targeted population model, over the
same stretch of mock data to try to recover the same set of lensed injections used in
the injection campaign to compare the performance of the TESLA and TESLA-X methods.
Table \ref{Table: TESLA-X Injection recovery} summarizes the findings. As expected, the TESLA
method successfully recovers more injections than the full template bank, with an
increase of $0.29\%$. However, we can see that the proposed TESLA-X method
outperforms the TESLA method. In particular, it recovers even more lensed injections
than the TESLA method, with an increase of $2.37\%$.
\begin{table}
\begin{center}
\begin{tabular}{c|c|c|c}
        Injections & General & TESLA & TESLA-X \\
        \hline
        Total & $7815$ & $7815$ & $7815$ \\
        Found & $6151$ & $6169$ &  $6297$ \\
        Found $\%$ change & - & $+0.29\%$ & $+2.37\%$ \\
\end{tabular}
\end{center}
\caption{
        \label{Table: TESLA-X Injection recovery}
        Number of injections found during the search of mock data using the general
        template bank, TESLA method, and TESLA-X method, respectively.
}
\end{table}
\begin{figure}
\centering
\includegraphics[width=\columnwidth]{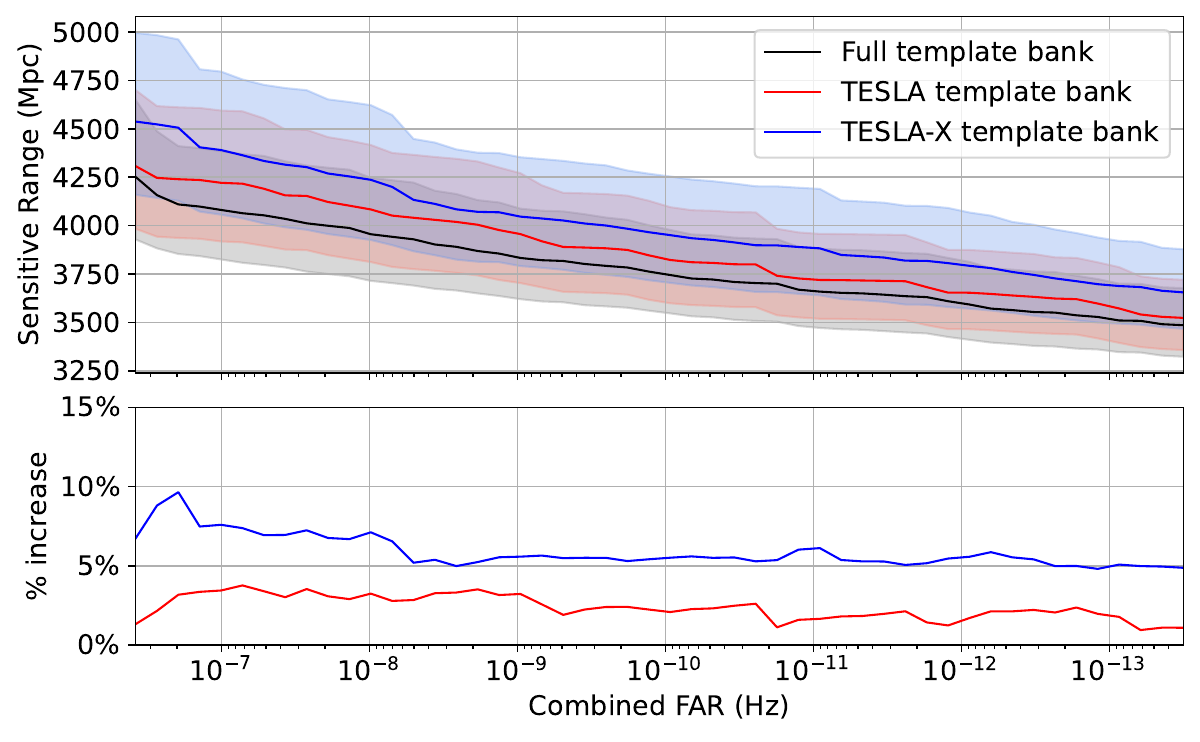}
\caption{
        \label{Fig: MS220508a range comparison}
        (Top panel) The sensitive range v.s. FAR for MS220508a-alike signals
        using the full template bank (black), TESLA template bank (red), and TESLA-X
        template bank (blue), respectively. The shaded band for each curve represents
        the corresponding $1$-sigma region. (Bottom panel) The corresponding percentage
        changes in sensitive range v.s. FAR for the different banks. We note that
        both the TESLA and TESLA-X bank improves in terms of sensitivity towards MS220508a-alike
        (lensed) signals, but the curve representing the TESLA-X bank (blue) is above that of the
        TESLA bank (red), meaning that the TESLA-X method outperforms the TESLA method. In particular,
        notice that the improvement is most significant in the region with FAR $\geq 10^{-7} \text{ Hz}$,
        which represents the ``sub-threshold'' region. This further proves that the TESLA-X method
        is better than the traditional TESLA method.
}
\index{figures}
\end{figure}
Furthermore, if we look at the sensitive range v.s. FAR for MS220508a-alike signals
using the full template bank, TESLA template bank, and TESLA-X template bank, respectively
(see Figure \ref{Fig: MS220508a range comparison}), we can see that the TESLA-X method gives
a bigger improvement in terms of sensitivity towards MS220508a-alike (lensed) signals than
the traditional TESLA method. In particular, we note that the improvement is most significant
in the region with FAR $\geq 10^{-7} \text{ Hz}$, which represents the ``sub-threshold'' region.
This means that TESLA-X outperforms TESLA in finding possible sub-threshold strongly lensed
gravitational waves.

%%%%%%%%%%%%%%%%%%%%%%%%%%%%%%%%%%%%%%%%%%%%%%%%%%%%%%%%%%%%%%%%%%%%%%%%%%%%%

\section{Concluding Remarks}\label{Section: Conclusion}
The detectability of strongly lensed gravitational wave (GW) signals can be compromised as
they may experience demagnification, falling below the detection threshold. In a previous
study, we introduced the TESLA method \citep{Li:2019osa} to extract strongly lensed sub-threshold
GW signals by constructing a target template bank based on the information from detected events.
This paper presents an enhanced approach called TESLA-X, which aims to improve the search
sensitivity for strongly lensed sub-threshold GW signals.\\
The TESLA-X methodology begins with generating simulated lensed injections using posterior
samples obtained from parameter estimation. These injections are then subjected to an injection
campaign using GstLAL, with a threshold set at a FAR of $3.385 \times 10^{-7}$ Hz, ensuring
statistical significance. Subsequently, the TESLA-X method utilizes the retrieved injections
to create a targeted population model and a densely distributed template bank. Finally, the
data is reprocessed to identify potential sub-threshold lensed candidates using
the new population model and template bank.\\
The targeted population model is constructed through several steps. Initially, a Gaussian kernel
density estimation (KDE) function is applied to the retrieved injections' source parameters $\vec{\gamma}$. This KDE function is then transformed into a probability density function,
representing the parameter space of our potential strong-lensed sub-threshold GW signals \citep{Heather}.
Additionally, we generate the targeted template bank by defining a boundary within the retrieved
injections. This process involves discarding templates from the original full template bank outside of a predefined boundary that encloses the target parameter space while keeping all templates within the boundary. This results in a densely distributed template bank targeting a smaller region.
Compared to the TESLA method, this refined template bank ensures the recovery of any
lensed counterparts with less than a $3\%$ SNR loss. Notably, the FARs and
likelihood values ($\mathcal{L}$) do not provide information about the likelihood of being lensed
counterparts to the target event. Instead, they serve as metrics to differentiate between noise
triggers and astrophysical signals.\\
In this study, we conducted a simulation campaign to evaluate and compare the search sensitivity
of three methods: the full template bank search, the TESLA method, and the TESLA-X
method. Our analysis utilized a stream of mock data, where Gaussian noise was introduced
along with a significant GW event labeled MS220508a. Our findings demonstrate that the
TESLA-X method outperforms the other approaches by successfully recovering more lensed injections specifically targeting MS220508a. Furthermore, we observed a maximum of
$\sim 10\%$ increase in the search sensitivity through an examination of the sensitive range versus
FAR plot. An increase in sensitive range means that we can see signals coming from further
sources, and equivalently, we can see weaker signals according to equation~\ref{Equation: TESLA-X-rho-distance}.
These results substantiate the superior performance of the TESLA-X method in enhancing the search
capabilities for gravitational lensing phenomena. TESLA-X still carries the same problem as TESLA
in terms of high computational cost (For reference, running TESLA/TESLA-X for a single super-threshold
event can take $\mathcal{O}(\mathrm{weeks})$.) for running the injection campaign for each target
super-threshold gravitational-wave event. As future work, we will explore ways to improve TESLA-X's computational efficiency. We will also look into possible parameters other than chirp
mass and effective spin to represent our interested population (See Appendix).

\section*{Acknowledgements}
The authors acknowledge the generous support from the National Science Foundation in the United States.
%The authors would also like to acknowledge Jonah Kanner and Bruce Allen for their useful suggestions.
%RKLL and TGFL would also like to gratefully acknowledge the support from the Croucher Foundation in Hong Kong.
%The work described in this paper was partially supported by a grant from the Research Grants Council of the Hong Kong (Project No. CUHK 14306218) and the Direct Grant for Research from the Research Committee of the Chinese University of Hong Kong.
%SS was supported in part by the LIGO Laboratory and in part by the Eberly Research Funds of Penn State, The Pennsylvania State University, University Park, PA 16802, USA.
AKYL would like to gratefully acknowledge the sup- port from the National Science Foundation through the Grants NSF PHY-1912594 and NSF PHY-2207758.
JCLC acknowledges support from the Villum Investigator program supported by the VILLUM Foundation (grant no. VIL37766) and the DNRF Chair program (grant no. DNRF162) by the Danish National Research Foundation.  
HF gratefully acknowledges the support of the Canadian Institute for Theoretical Astrophysics (CITA) National Fellowship and the Japan Society for the Promotion of Science (JSPS) Postdoctoral Fellowships for Research in Japan. 
JME is supported by the European Union’s Horizon 2020 research and innovation program under the Marie Sklodowska-Curie grant agreement No. 847523 INTERACTIONS, and by VILLUM FONDEN (grant no. 53101 and 37766).\\
The authors are also grateful for computational resources provided by the LIGO Laboratory and supported by National Science Foundation Grants PHY-0757058 and PHY-0823459.
This research has made use of data, software and/or web tools obtained from the Gravitational Wave Open Science Center (https://www.gw-openscience.org) \citep{2015JPhCS.610a2021V}, a service of LIGO Laboratory, the LIGO Scientific Collaboration and the Virgo Collaboration. LIGO was constructed by the California Institute of Technology and Massachusetts Institute of Technology with funding from the National Science Foundation and operates under cooperative agreement PHY-0757058. HF is supported by a CITA National Fellowship.
Virgo is funded by the French Centre National de Recherche Scientifique (CNRS), the Italian Istituto Nazionale della Fisica Nucleare (INFN) and the Dutch Nikhef, with contributions by Polish and Hungarian institutes.
This paper carries LIGO Document Number LIGO-\DocumentID{}.

%%%%%%%%%%%%%%%%%%%%%%%%%%%%%

\appendix
\section{Choosing a suitable pair of parameters to construct the Gaussian KDE}

In the main text, we construct the Gaussian KDE in the chirp mass $\mathcal{M}_c$ - effective spin $\chi_\text{eff}$
parameter space. It has also been suggested that the component masses $m_1$-$m_2$ parameter space
may be a better choice. In this appendix, we present some preliminary investigation results for this question.

\subsection{Re-doing the TESLA-X analysis for MS220508a with the Gaussian KDE constructed in the $m_1$-$m_2$ space}
In this subsection, we redo the TESLA-X analysis with the Gaussian KDE constructed in the component masses $m_1$-$m_2$ parameter space
instead of the chirp mass $\mathcal{M}_c$ - effective spin $\chi_\text{eff}$ parameter space.

\subsubsection{Constructing the TESLA-X bank and population model}
\begin{figure}
\centering
\includegraphics[width=\columnwidth]{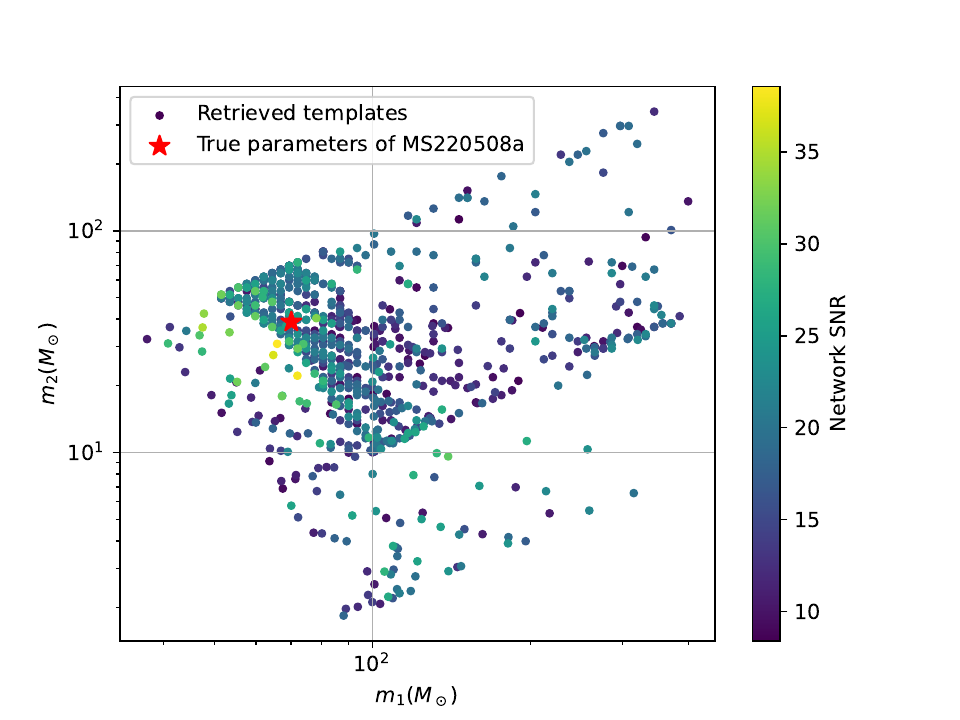}
\caption{
        Rung-up templates by the simulated lensed injections of MS220508a in the injection
        campaign, plotted on the component masses $m_1$-$m_2$ parameter space.
        The colors of the markers represent the network SNRs of the associated injections.
}
\label{Fig: MS220508a recovered templates m1m2}
\index{figures}
\end{figure}

\begin{figure}
\centering
\includegraphics[width=\columnwidth]{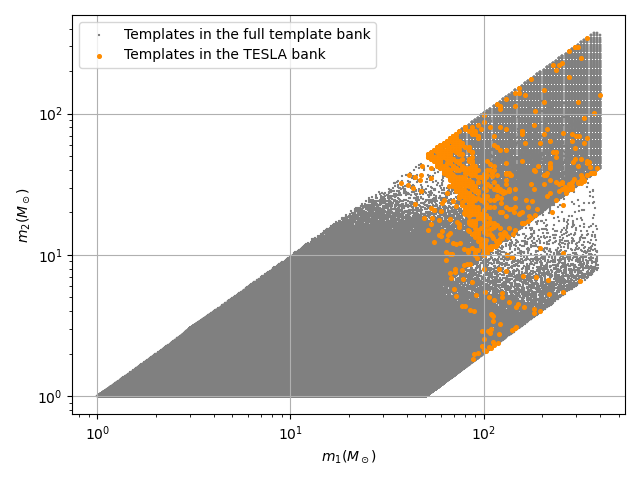}
\caption{
        The templates in the full template bank (in grey) and in the TESLA reduced template
        bank (in orange), plotted in the component masses $m_1$-$m_2$ space.
}
\label{Fig: MS220508a TESLA template bank m1m2}
\index{figures}
\end{figure}

\begin{figure}
\centering
\includegraphics[width=\columnwidth]{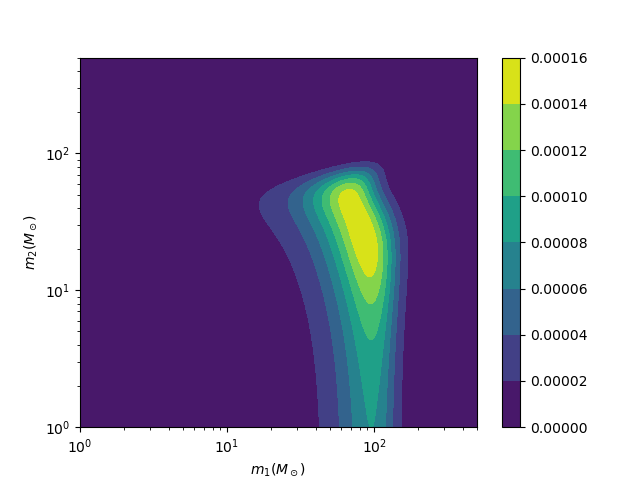}
\caption{
        The contour map of the Gaussian Kernel Density Estimation function obtained for
        MS220508a in the $m_1$ - $m_2$ space. The colors represent the probability density
        estimated by the KDE function.
}
\label{Fig: MS220508a Gaussian KDE m1m2}
\index{figures}
\end{figure}

\begin{figure}
\centering
\includegraphics[width=\columnwidth]{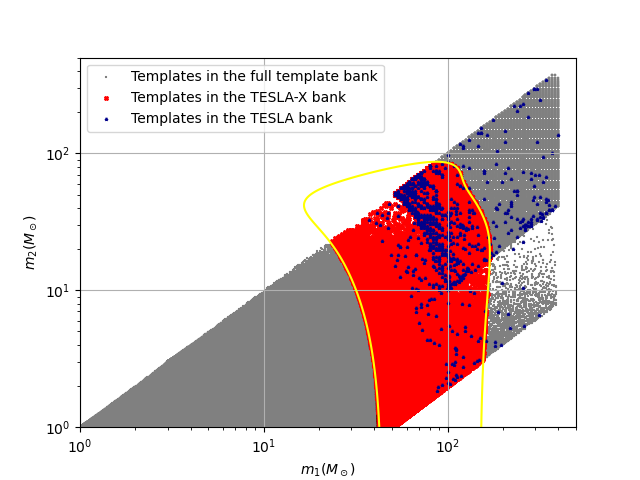}
\caption{
        The templates in the full template bank (in grey) and in the ``TESLA-X (component mass) bank'' (in red),
        plotted in the $m_1$-$m_2$ space. The ``TESLA-X (component mass) bank'' is constructed by
        keeping templates that fall within the lowest contour level from the Gaussian KDE function shown in
        Figure \ref{Fig: MS220508a Gaussian KDE m1m2}, i.e. $0.00002$, plotted as a yellow curve in this
        figure. For easy comparison, we also plot the templates in the TESLA reduced template bank in the
        same figure (in blue).
}
\label{Fig: MS220508a TESLA-X bank m1m2}
\index{figures}
\end{figure}

We start with the injection run results we obtained previously for the same event.
First, Figure \ref{Fig: MS220508a recovered templates m1m2} plots the rung-up templates on the
component masses $m_1$-$m_2$ parameter space, with the colors of the markers being the network
SNRs of the associated injections.
Note that although we are working in a different parameter space, templates in the TESLA
template bank will remain unchanged because we are simply keeping all the templates that
are rung up by the injections during the injection campaign to form the TESLA bank. For
completeness, we also plot the \textbf{same} TESLA bank in the main text for MS220508a on the
component masses $m_1$-$m_2$ parameter space in Figure \ref{Fig: MS220508a TESLA template bank m1m2}
As before, templates in the full template bank are plotted in grey, and those in the TESLA template bank are
plotted in orange.
For TESLA-X, we use the rung-up templates in the injection campaign to construct a
Gaussian Kernel Density Estimation (KDE) function $\text{KDE}(\vec{\gamma})$ with the
associated injections' network SNRs as the weights. As before, $\gamma$ represents the template
parameters used to label the templates. Here, $\vec\gamma = \{m_1, m_2\}$.
Then, we evaluate the probability density at each template point in the full template bank,
and plot the results on a contour map. Figure \ref{Fig: MS220508a Gaussian KDE m1m2} shows the
contour map for the Gaussian KDE function estimated for MS220508a in the $m_1$-$m_2$ space.
We again take the lowest contour level ($0.00002$) as a boundary, and use all the templates that fall within this
boundary to construct a TESLA-X targeted template bank in the $m_1$-$m_2$ space. For simplicity,
we will call the TESLA-X bank created based on the KDE constructed in the $m_1$-$m_2$ space as the
``TESLA-X (component mass) bank'', and the TESLA-X bank in the main text that is created based on the KDE constructed
in the $\mathcal{M}_c$ - $\chi_\text{eff}$ space as the ``TESLA-X (main) bank''.
Figure \ref{Fig: MS220508a TESLA-X bank m1m2} shows the templates in the full template bank (in grey)
and in the ``TESLA-X (component mass) bank'' (in red) in the $m_1$-$m_2$ space. The yellow curve marks the
lowest contour level we obtained for the estimated Gaussian KDE function we showed in
Figure \ref{Fig: MS220508a Gaussian KDE m1m2}. There are $159043$ templates in the ``TESLA-X (component mass) bank''.
Similar to the main text, we use the estimated KDE function for MS220508a and the ``TESLA-X (component mass) bank''
to construct a targeted population model, following the same steps detailed in \ref{Section: TESLA-X}
and \citep{Heather}.

\subsubsection{Trying to recover the lensed injections with the ``TESLA-X (main) bank''  and ``TESLA-X (component mass) bank''}
To compare the effectiveness of the ``TESLA-X (main) bank''  and ``TESLA-X (component mass) bank'' for the case of MS220508a,
we perform an additional search using the ``TESLA-X (component mass) bank'' together with the targeted population model
over the same stretch of mock data described in the main text to try to recover the same set of lensed injections
used in the injection campaign.
Table \ref{Table: TESLA-X Injection recovery m1m2} summarizes the findings.
We can see that ``TESLA-X (component mass)'' is performing even worse
than the full template bank, with a reduction in recovered lensed injections by
$-0.3\%$. This should not be surprising, however, given that the `` TESLA-X (component mass) bank''
has more templates than the ``TESLA-X (main) bank'' (almost six times more templates).
This results in a much higher trials factor and, hence, a larger noise background,
making it harder to recover the simulated lensed injections.
\begin{table}
\begin{center}
\begin{tabular}{c|c|c|c}
        Injections & General & TESLA-X & TESLA-X \\
        & & (main) & (component mass) \\
        \hline
        Total & $7815$ & $7815$ & $7815$ \\
        Found & $6151$ & $6297$ & $6132$\\
        Found $\%$ change & - & $+2.37\%$ & $-0.3\%$\\
\end{tabular}
\end{center}
\caption{
        \label{Table: TESLA-X Injection recovery m1m2}
        Number of injections found during the search of mock data using the general
        template bank, ``TESLA-X (main) bank'' and ``TESLA-X (component mass) bank'' respectively.
}
\end{table}
\begin{figure}
\centering
\includegraphics[width=\columnwidth]{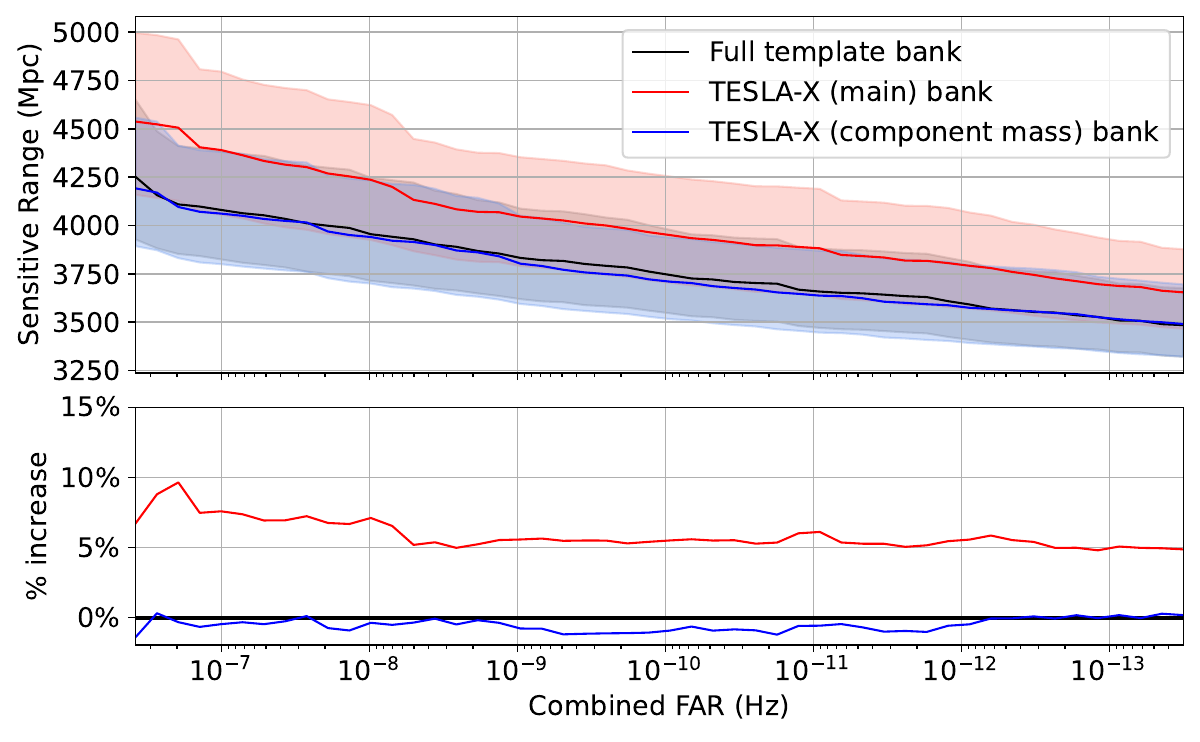}
\caption{
        \label{Fig: MS220508a range comparison m1m2}
        (Top panel) The sensitive range v.s. FAR for MS220508a-alike signals
        using the full template bank (black), ``TESLA-X (main) bank'' (red), and ``TESLA-X (component mass)
        bank'' (blue) respectively. The shaded band for each curve represents
        the corresponding $1$-sigma region. (Bottom panel) The corresponding percentage
        changes in sensitive range v.s. FAR for the different banks. We note that
        while the ``TESLA-X (main) bank'' improves in terms of sensitivity towards MS220508a-alike
        (lensed) signals, ``TESLA-X (component mass) bank'' is performing worse than the full
        template bank, mainly due to the significant increase in trials factors caused by
        the large number of templates.
}
\index{figures}
\end{figure}
Furthermore, if we look at the sensitive range v.s. FAR for MS220508a-alike signals
using the full template bank, ``TESLA-X (main) bank'' and ``TESLA-X (component mass) bank'' respectively
(see Figure \ref{Fig: MS220508a range comparison m1m2}), we can see that the ``TESLA-X (component mass)
bank'' is performing even worse than the full template bank. Therefore, in the case of
MS220508a, chirp mass $\mathcal{M}_c$ - effective spin $\chi_\text{eff}$ would be the better pair
of parameters to construct the Gaussian KDE in the TESLA-X method.

\subsection{Analyzing another mock event MS220510ae with the TESLA-X method}
\begin{figure}
\centering
\includegraphics[width=\columnwidth]{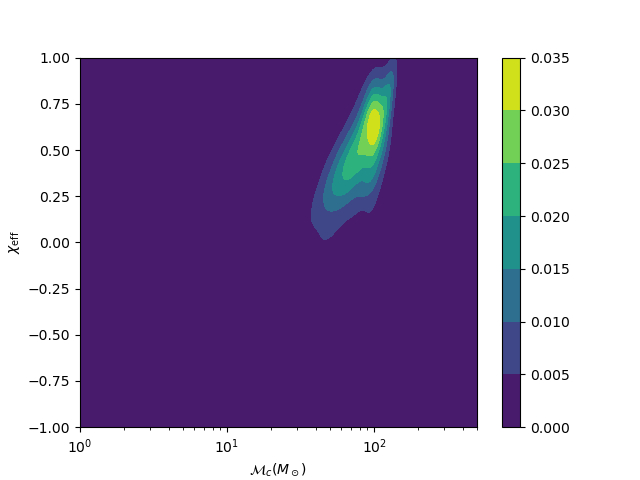}
\caption{
        The contour map of the Gaussian Kernel Density Estimation function obtained for
        MS220510ae in the $\mathcal{M}_c$ - $\chi_\text{eff}$ space. The colors represent the probability density
        estimated by the KDE function.
}
\label{Fig: MS220510ae Gaussian KDE}
\index{figures}
\end{figure}

\begin{figure}
\centering
\includegraphics[width=\columnwidth]{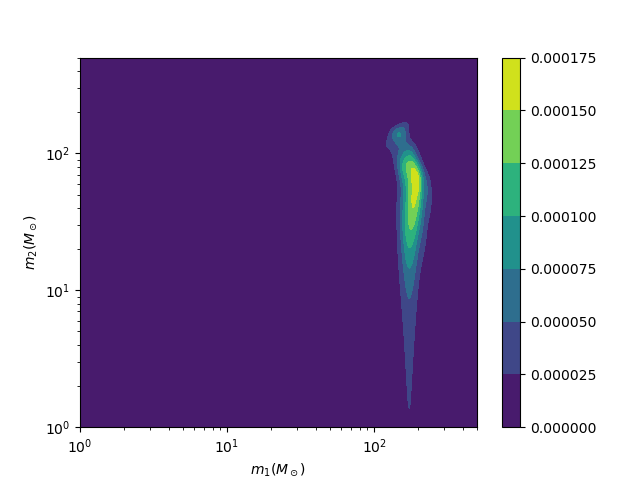}
\caption{
        The contour map of the Gaussian Kernel Density Estimation function obtained for
        MS220510ae in the $m_1$ - $m_2$ space. The colors represent the probability density
        estimated by the KDE function.
}
\label{Fig: MS220510ae Gaussian KDE m1m2}
\index{figures}
\end{figure}

\begin{figure}
\centering
\includegraphics[width=\columnwidth]{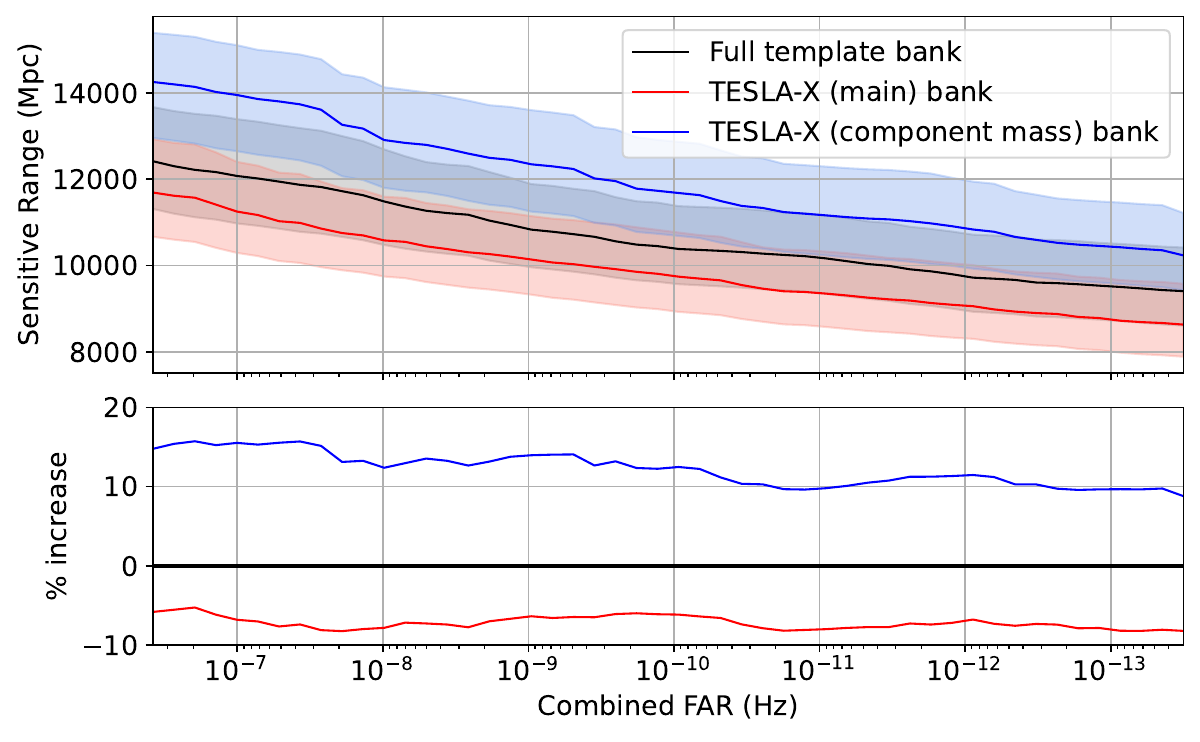}
\caption{
        \label{Fig: MS220510ae range comparison m1m2}
        (Top panel) The sensitive range v.s. FAR for MS220510ae-alike signals
        using the full template bank (black), ``TESLA-X (main) bank'' (red), and ``TESLA-X (component mass)
        bank'' (blue) respectively. The shaded band for each curve represents
        the corresponding $1$-sigma region. (Bottom panel) The corresponding percentage
        changes in sensitive range v.s. FAR for the different banks. We note that, unlike the case of MS220508a in the main text,
        the ``TESLA-X (component mass) bank'' improves in terms of sensitivity towards MS220510ae-alike
        (lensed) signals, ``TESLA-X (main) bank'' performs worse than the full
        template bank. This is mainly because spins are typically poorly measured for high chirp mass systems.
}
\index{figures}
\end{figure}

MS220508a is a relatively low chirp mass event ($\sim 45.0 M_\odot$). It will be
useful to redo the above investigation about the choice of parameters to construct the Gaussian KDE
in the TESLA-X method with a different mock event that lives in a different region in the parameter space.\\
In this subsection, we repeat the analysis in the previous subsection with another mock superthreshold
gravitational-wave event MS220510ae. Settings for the mock data are identical to those in the main text.
Details about the source parameters of the injected gravitational wave are listed in
Table \ref{Table: MS220510ae}.
\begin{table}
\begin{center}
%\begin{adjustbox}{minipage=1.0\columnwidth, center}
\begin{tabular}{c|c}
        Properties & Injected super-threshold signal \\
        \hline
        UTC time & May 10 2022 $18:38:17$\\
        GPS time & $1336243115.828$ \\
        Distance (Mpc) & $10066.811$  \\
        \hline
        Primary mass $m_1^\text{det}$ & {$166.90 M_\odot$}\\
        Secondary mass $m_2^\text{det}$ & {$117.04 M_\odot$}\\
        Dimensionless spins & {$\chi_{1x}=0.635$}, {$\chi_{1y}=0.233$}, {$\chi_{1z}=0.660$},\\
        & {$\chi_{2x}=0.006$}, {$\chi_{2y}=0.087$}, {$\chi_{2z}=0.029$},\\
        Right ascension $\alpha$ & {$5.69$} \\
        Declination $\delta$ & {$1.45$} \\
        Inclination $\iota$ & {$5.99$}\\
        Polarization $\Psi$ & {$2.92$}\\
        Waveform & IMRPhenomXPHMpseudoFourPN \\
\end{tabular}
%\end{adjustbox}
\end{center}
\caption{
        \label{Table: MS220510ae}
        Information of the injected super-threshold gravitational-wave signal MS220510ae
        in the simulation campaign. All properties reported here are measured in the detector frame.
}
\end{table}
We follow the same steps in the main text to perform Bayesian parameter estimation on MS220510ae,
and use the posterior samples obtained to generate a set of lensed injections. We then conduct an
injection campaign using the full template bank to search for the lensed injections. The injection run
results are then used to construct the ``TESLA-X (main) bank'' and ``TESLA-X (component mass) bank'', and their
corresponding targeted population models.
Out of the $8099$ lensed injections, the full template bank manages to recover $5047$.
For simplicity, we only show the Gaussian KDE constructed for the ``TESLA-X (main) bank'' (in the $\mathcal{M}_c$-$\chi_\text{eff}$ space)
and for the ``TESLA-X (component mass) bank'' (in the $m_1$-$m_2$ space) in Figure \ref{Fig: MS220510ae Gaussian KDE} and \ref{Fig: MS220510ae Gaussian KDE m1m2} respectively.
As before, we construct the ``TESLA-X (main) bank'' and ``TESLA-X (component mass) bank'' by keeping all templates that fall
within the boundary defined by the least significant contour in the respective Gaussian KDE constructed.
For the case of MS220510ae, the ``TESLA-X (main) bank'' contains $15655$ templates, and the ``TESLA-X (component mass) bank'' contains $21676$ templates.
Notice that the ``TESLA-X (component mass) bank'' for MS220510ae, unlike in the case for MS220508a, is larger than the ``TESLA-X (main) bank''
by merely $\sim 5000$ templates.\\
We then use the two TESLA-X banks and their respective population models to try recovering the same set of lensed injections
used in the injection campaign.
Table \ref{Table: TESLA-X MS220510ae Injection recovery} summarizes the findings.
Completely different from the case of MS220508a in the main text,
we can see that ``TESLA-X (component mass)'' is performing even better
than the full template bank, while the ``TESLA-X (main) bank'' is performing worse than
the full template bank.
This is mainly because spins are poorly measured in general for high-mass systems,
and hence the ``TESLA-X (main) bank'', with inaccurate constraint in the spin
dimension, are more likely to lose signals than the ``TESLA-X (component mass) bank'', which
does not impose any constraints on spins. For completeness, we also show the sensitive range v.s. FAR for MS220510ae-alike signals
using the full tempalte bank, ``TESLA-X (main) bank'' and ``TESLA-X (component mass) bank'' respectively in Figure \ref{Fig: MS220510ae range comparison m1m2}.
\begin{table}
\begin{center}
\begin{tabular}{c|c|c|c}
        Injections & General & TESLA-X & TESLA-X \\
        & & (main) & (component mass) \\
        \hline
        Total & $8099$ & $8099$ & $8099$ \\
        Found & $5047$ & $4715$ & $5886$\\
        Found $\%$ change & - & $-6.58\%$ & $16.6\%$\\
\end{tabular}
\end{center}
\caption{
        \label{Table: TESLA-X MS220510ae Injection recovery}
        Number of injections found during the search of mock data using the general
        template bank, ``TESLA-X (main) bank'' and ``TESLA-X (component mass) bank'' respectively for the mock event MS220510ae.
}
\end{table}

\subsection{Conclusion}
From the two investigations, we note that ``chirp mass - effective spins'' is a better parameter pair to be
used in constructing the Gaussian KDE for the TESLA-X method, if the target event is a low chirp mass system.
On the other hand, component masses ``mass $1$ - mass $2$'' will be a better parameter pair to be
used in constructing the Gaussian KDE for the TESLA-X method, if the target event is a high chirp mass system.
We note that these findings are entirely expected: While chirp masses and spins are well-measured in low-mass systems, they are not well-measured in high-mass systems since their signals have much fewer in-band cycles.
As future work, we will investigate how we can improve the construction of the Gaussian KDE functions for the TESLA-X method. For instance, we can see that the Gaussian KDEs can exceed the physical boundary of the original template bank (See the yellow boundary in \ref{Fig: MS220508a TESLA-X bank m1m2}). Regions near the physical boundary of the original template bank may require special care. We may also explore the idea of first constructing and applying a separate KDE to ``flatten" the  TESLA-X bank (i.e., to construct a uniform mass model) before applying the Gaussian KDE representing the targeted population model.

\bibliographystyle{mnras}
\bibliography{citations}
\end{document}